\documentclass[aps,nofootinbib,twocolumn,showpacs,superscriptaddress,amssymb]{revtex4-1}
\date{\today}
\usepackage{graphicx}
\usepackage{color}
\begin{document}

\title{On the Validity of the Geiger-Nuttall Alpha-Decay Law and its Microscopic Basis}

\author{C. Qi}
\email{chongq@kth.se}
\affiliation{KTH, Alba Nova University Center,
SE-10691 Stockholm, Sweden}

\author{A. N. Andreyev}
\affiliation{Department of Physics, University of York, YO10 5DD, United Kingdom}
\affiliation{Advanced Science Research Center, Japan Atomic Energy Agency (JAEA),
Tokai-mura,  319-1195, Japan}

\author{M. Huyse}
\affiliation{KU Leuven, Instituut voor Kern-en Stralingsfysica, 3001 Leuven, Belgium}

\author{R. J. Liotta}
\affiliation{KTH, Alba Nova University Center,
SE-10691 Stockholm, Sweden}

\author{P. Van Duppen}
\affiliation{KU Leuven, Instituut voor Kern-en Stralingsfysica, 3001 Leuven, Belgium}

\author{R. Wyss}
\affiliation{KTH, Alba Nova University Center, SE-10691 Stockholm,
Sweden}
\begin{abstract}
 The Geiger-Nuttall (GN) law relates the partial $\alpha$-decay half-life
with the energy of the escaping $\alpha$ particle and
contains for every isotopic chain  two experimentally determined coefficients. The expression is supported by several phenomenological approaches, however its coefficients lack a fully microscopic basis.  In this paper we will show that: 1) the empirical coefficients that appear in the
GN law 
have a deep physical meaning and 2) the GN law is successful within the restricted experimental data sets available so far, but  is not valid in general. 
We will show that, when the dependence of logarithm values of the $\alpha$ formation probability on the neutron 
number is not linear or constant, the GN law is broken. For the $\alpha$ decay of neutron-deficient nucleus $^{186}$Po, the difference between the experimental 
half-life and that predicted by the GN Law is as large as one order of magnitude. 

\end{abstract}

\maketitle

\begin{figure}
\vspace{0.3cm}
\begin{center}
\includegraphics[scale=0.43]{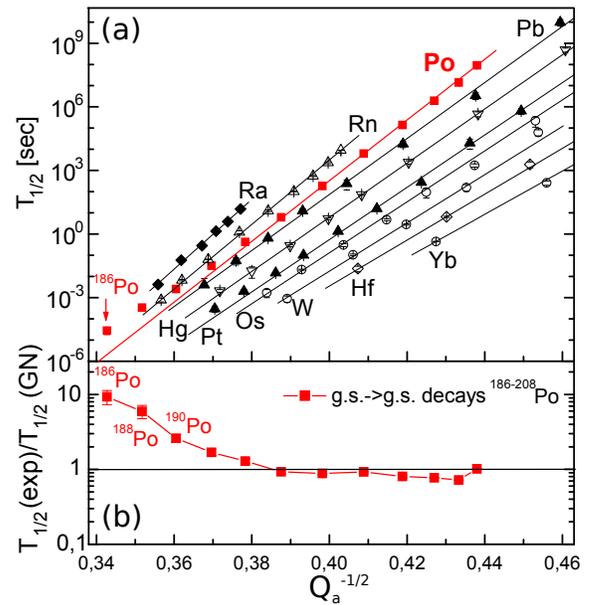}
\end{center}
\caption{\label{gn} (a): The logarithms of experimental
partial $\alpha$-decay half-lives (in
sec)~\cite{aud03,nndc,And12,Van03}  for the even-even Yb-Ra nuclei with neutron number $N<126$ as a function
of $Q_\alpha^{-1/2}$
(in MeV$^{-1/2}$). The straight lines show the description of the GN law with A and B values fitted for each isotopic chain. (b): The deviation of
the experimental $\alpha$-decay half-lives from those predicted 
by the GN law for the light Po isotopes. }
\end{figure}

One landmark in modern physics, shaping the developments leading to Quantum
Mechanics, was the formulation of the empirical Geiger-Nuttall (GN) law in 1911 
\cite{gn}.  
According to the GN law as
formulated in Ref.  \cite{gn},  the $\alpha$ decay partial half-life 
$T_{1/2}$ is given by, 
\begin{equation}\label{gn-1} \log_{10} T_{1/2} =
A(Z)Q_{\alpha}^{-1/2}+B(Z), 
\end{equation} 
where $Q_{\alpha}$ is the total energy of the $\alpha$ decay process ($\alpha$-decay $Q$ 
value) and $A(Z)$ and $B(Z)$ are the coefficients which are
determined by fitting experimental data for each isotopic chain. The GN law has been verified in long isotopic chains and no strong deviations have been observed: It is extremely successful and is considered to be generally valid. 
 Recently the amount of $\alpha$ decay data in
heavy and superheavy nuclei has greatly increased
\cite{aud03,nndc,And10,Pf12,Van03,And12} and  the GN law is still fulfilled,
reproducing most experimental data within a factor 2$\sim$3, as seen in Fig. \ref{gn}a for the Yb-Ra region (apart from the Po chain, as will be discussed in this paper).  The coefficients
$A$ and $B$ give rise to
different GN lines for each isotope series (Fig. \ref{gn}a).  The coefficients change for each isotopic chain which crosses the magic numbers, e.g. $N=126$~\cite{Buck90}.

The greatest challenge was thus to understand how the $\alpha$ particle could leave the mother
nucleus without any external agent disturbing it.
The first successful theoretical explanation was given by Gamow \cite{Gam29} and 
independently by Condon and Gurney \cite{con28}, who explained $\alpha$ decay as the penetration (tunneling) through the Coulomb barrier, leading to the $Q_{\alpha}^{-1/2}$ dependence of Eq. (1). This was a great revolution in physics and confirmed the
probabilistic interpretation of Quantum Mechanics. 
Besides its pioneering role in the development of quantum theory, the $\alpha$ decay also broadens our understanding of the quantum tunneling process of other composite objects. This is a general physical process that can be found in other fields including condensed matter, molecular physics and astrophysics \cite{Bala98,Raz03,Lem09}.

\begin{figure}
\includegraphics[scale=0.35]{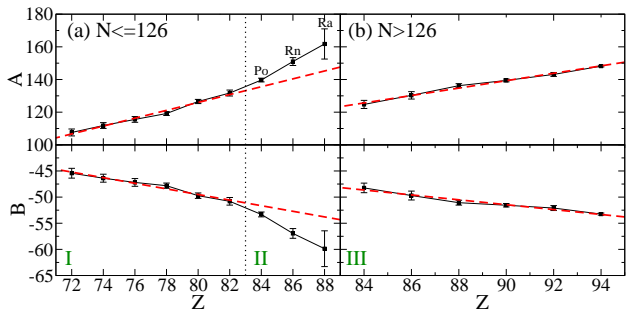}
\caption{\label{abngt126}(a): The coefficients $A(Z)$ and $B(Z)$ for even-even 
nuclei in regions I ($Z\leq82$) and II
($Z>82$)  with 
$N\leq126$. The red dashed lines are fitted only for the data from region I with $Z\leq82$ giving $A(Z)=2.41Z-66.7$ and $B(Z) = -0.54Z-6.61$. (b): 
Same as (a) but
for nuclei in region III ($Z>82$), i.e., polonium to plutonium isotopes
with neutron numbers $N>126$. Again, the red dashed lines are determined by a fitting 
procedure, which gives $A(Z)=2.27Z-65.0$ and $B(Z) = -0.47Z-9.36$.}
\end{figure}

The remaining challenge is to identify the microscopic basis of the GN coefficients $A$ and $B$. 
Thomas  provided the first attempt of a microscopic theory 
\cite{Tho54}. His expression for the
half life considered the probability that the four nucleons, which 
eventually constitute the $\alpha$ particle, get clustered at a certain 
distance on
the nuclear surface. 
In this paper we will apply Thomas's expression to probe the general validity of the GN law, explain the $Z$ dependence of the constants and their microscopic origin. 

Guided by recent experimental findings, we divide the $\alpha$-decaying nuclei into four
regions (see Fig. \ref{abngt126}): \\
I) $N\leq126$, $Z\leq82$; \\
II) $N\leq126$, $Z>82$;\\
III) $N>126$, $Z>82$;\\
IV) $N>126$, $Z\leq82$.\\
Except for $^{210}$Pb, $\alpha$ decay has not yet been observed for nuclei in region IV. Fig. \ref{abngt126} shows that
both coefficients $A$ and $B$ are linearly 
dependent upon $Z$  for regions I and III, however with different coefficients. This 
was initially reported in Ref. 
\cite{Vio66} and attributed to the crossing of the $N=126$ neutron shell. The recent extension of the available data  also shows that,
when crossing the $Z=82$ shell gap (from region I to region II), another set of coefficients, strongly deviating from the values in region I and III, is needed.
Further, as seen in Fig. 1b, recent $\alpha$-decay experiments at SHIP in GSI (Darmstadt, Germany) to study the neutron-deficient Po 
isotopes \cite{And12} show a significant and gradually increasing
deviation from the GN law using the coefficients as reported in Fig. 2. In $^{186}$Po, the difference between the experimental 
half-life and that predicted by the GN Law is as large as one order of magnitude. 
Such a strong deviation has not been seen before.

In order to understand 
the three issues mentioned above we go
through the derivations of Refs. \cite{Qi09a,Qi09b} where a generalization of the GN 
law was found. This generalization  holds for all isotopic chains and all cluster
radioactivities.
According to Ref. \cite{Tho54}, the $\alpha$-decay half-life can be 
written as 
\begin{equation}\label{life}
T_{1/2}= \frac{\ln2}{\nu} \left|
\frac{H_l^+(\chi,\rho)}{RF_{\alpha}(R)} \right|^2,
\end{equation}
where $\nu$ is the velocity of the emitted $\alpha$ particle which
carries an angular momentum $l$. As only ground-state to ground-state $\alpha$ decays of even-even nuclei are considered here, $l$ is equal to 0 in all cases.
$R$ is a distance around the
nuclear surface where the wave function describing the cluster in
the mother nucleus is matched with the outgoing cluster+daughter
wave function. 
$H^+$ is the Coulomb-Hankel function with $\rho=\mu\nu R/\hbar$ and
$\chi = 4Ze^2/\hbar\nu$, the Coulomb parameter, where $\mu$ is the reduced mass 
and $Z$ is the charge number of the daughter
nucleus. The quantity $F_{\alpha}(R)$ is the formation
amplitude of the $\alpha$ cluster at distance $R$. 
Introducing the quantities $\chi' = 2Z\sqrt{A_{\alpha d}/Q_{\alpha}}$ and 
$\rho' = \sqrt{2 A_{\alpha d} Z(A_d^{1/3}+4^{1/3})}$ where 
$A_{\alpha d}=4 A_d/(4+A_d)$,
one gets, after imposing the condition of the half life being independent on $R$
\cite{Qi09a}
\begin{eqnarray}\label{gn-2}
\log T_{1/2}=a\chi' + b\rho' + c~~~~\\
\nonumber=2aZ/\sqrt{A_{\alpha d}}Q_{\alpha}^{-1/2}+b\sqrt{2 A_{\alpha d} Z(A_d^{1/3}+4^{1/3})}+c,
\end{eqnarray}
where $a$, $b$ and $c$ are constant parameters which only depend upon local
variations of the formation probability. They are determined by fitting 
experimental data \cite{Qi09a}. 

The reason why these parameters are practically constant is that,  when going 
from one isotope to another, 
the $\alpha$-particle formation
probability usually varies much less
than the penetrability. On the logarithm scale of 
the GN law the differences in the formation probabilities are
usually small fluctuations along the straight lines predicted
by that law. In other words, the constancy of the parameters $a$, $b$ and $c$ 
is a consequence of the
smooth variation in the nuclear structure that is often found
when going from a nucleus to its neighbors. This
is also the reason why, for example, the BCS approximation
works so well in many regions of nuclei.

\begin{figure}
\begin{center}
\includegraphics[scale=0.45]{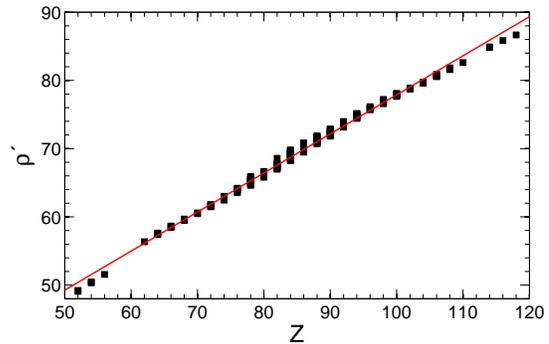}
\end{center}
\caption{\label{abz3}(Color online) Calculated values of $\rho'$ for observed even-even 
$\alpha$ emitters  as a function of $Z$. The linear fit to the values is shown by the solid line.}
\end{figure}

The term $a\chi'$ takes into account the tunneling through the Coulomb barrier, while $b\rho'+ c$,
which does not depend upon $Q_{\alpha}$, includes effects
induced by the clusterization in the mother nucleus \cite{Qi09a}. By comparing equations (1) and (3), a correspondence between the coefficients  $A(Z)$ and $B(Z)$ and the expressions $a\chi'$ and $b\rho'+ c$ respectively can be deduced and 
the meaning of the coefficients can be unfolded. 
The observed linear dependency of $A(Z)$ upon $Z$ is substantiated by this representation.
The observed negative values for $B(Z)$ are understood as both terms $b$ and $c$ are negative \cite{Qi09a}. The linear dependence upon $Z$ of $B(Z)$ seems to be in conflict with the $Z^{1/2}$ dependence of the term $\rho'$. However 
 for nuclei with known $\alpha$-decay half-lives for which the
GN law has so far been applied, $\rho'$ is practically
a linear function of $Z$, as seen in Fig. \ref{abz3}.

\begin{figure}
\begin{center}
\includegraphics[scale=0.3]{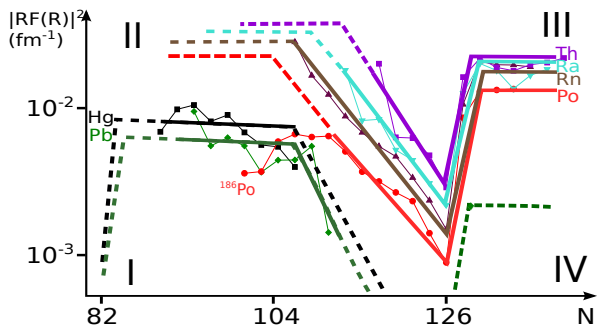}
\end{center}
\vspace{-0.2cm}
\caption{\label{gene} A  pictorial representation of the generic form of the evolution of the $\alpha$ formation probabilities $|RF(R)|^2$. Thick solid  lines are for isotopes, where experimental data are available and dashed lines are extapolations to the regions with the yet unavailable data. The  experimental data as cited in \cite{And12} are shown by points, connected by thin lines, to guide the eye. 
}
\end{figure}

The need for a different linear $Z$ dependence of the coefficients $A$ and $B$ in the four regions of the nuclear chart (see Fig. 2) will now be addressed. A generic form for the evolution of the alpha formation probabilities was proposed in \cite{And12}. It was based on  experimental values \cite{nndc,And10,Van03,And12} and 
calculations performed within the framework of the seniority scheme. This generic form is shown in Fig. 4 for selected isotopic chains.  Three distinct features can be extracted from this schematic representation.

The experimental $\alpha$ formation probabilities of most known $\alpha$ emitters in regions I and III are nearly constant as a function of neutron number 
(or more exactly, weakly linearly dependent on $\rho'$, as seen in Fig. 1 of 
Ref. \cite{Qi10}). For those nuclei, the GN law is indeed expected to be valid and $A(Z)$ and $B(Z)$ follow a linear behavior as a function of $Z$ (see Figs. 1 \& 3 and Eq. (9) in Ref. \cite{Qi09b}).

\begin{figure}
\begin{center}
\includegraphics[scale=0.35]{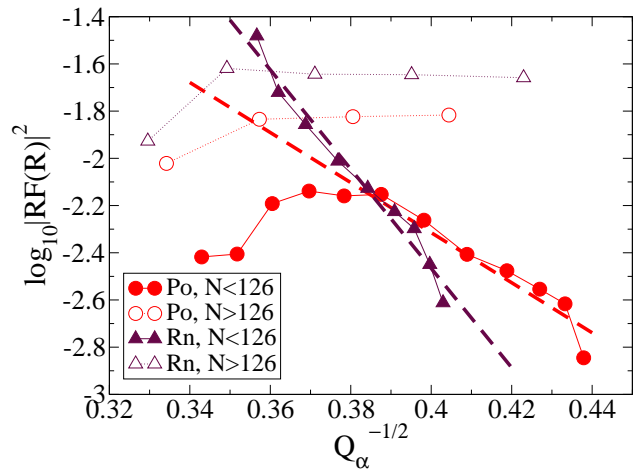}
\end{center}
\vspace{-0.2cm}
\caption{\label{fvsq} $log_{10}|RF(R)|^2$ for Po (circle) and Rn (triangle) isotopes in region II with $N<126$ (closed symbols) and region III with $N>126$ (open symbols) as a function of $Q_{\alpha}^{-1/2}$. The dashed lines are to guide the eye.}
\end{figure}

Approaching the $N = 126$ shell closure with increasing neutron number, a strong, exponential decrease of the  formation
probability is observed (see Fig. 4, in region II and the discussion in Ref. \cite{And12}). It is striking that in spite of a variation of $|RF_{\alpha}(R)|^2$ over one order of magnitude, the GN law and the $A(Z)$ and $B(Z)$ linear dependence upon $Z$ are still valid. This has no real physical meaning, but is a consequence of the specific dependence of the $|RF_{\alpha}(R)|^2$ on $Q_{\alpha}$. 
The $Q_{\alpha}$ (as well as $Q_{\alpha}^{-1/2}$) values exhibit a quasi linear pattern as a function of rising neutron number when  approaching the $N=126$ shell closure. Therefore $log_{10}|RF(R)|^2$ and thus  $log_{10}(T_{1/2})$ will still depend linearly on $Q_{\alpha}^{-1/2}$.  As examples, in Fig. \ref{fvsq}  the logs of the $\alpha$ formation probabilities $|RF(R)|^2$ for polonium and radon isotopes in regions II \& III are shown as a function of $Q_{\alpha}^{-1/2}$. In comparison with those in region III for which the formation probabilities are nearly constant or only weakly depend on $Q_{\alpha}^{-1/2}$, the data in region II show an exponential dependence. The other isotopic chains in region II show a similar linearly decreasing behavior  of $log_{10}|RF(R)|^2$ as a function of $Q_{\alpha}^{-1/2}$, as indicated by the red-dashed lines in the figure, however with different slopes. As a result, the GN law remains valid for isotopic chains in region II, but the corresponding values of $A$ and $|B|$ will increase with $Z$ beyond the trend observed in region I and III (see Fig. 2).

For the polonium isotopic chain with $N<126$, the linear behavior of $log_{10}|RF(R)|^2$ breaks down below $^{196}$Po  ($N=112$, corresponding to $Q_{\alpha}^{-1/2}=0.39$ in Fig. 5). 
This explains why the GN law is broken in the light polonium isotopes of
Fig. \ref{gn}b. This violation of the GN law, observed here for the
first time, is induced by the strong suppression of the $\alpha$ formation 
probability due to the fact that the deformations and configurations of the ground states of the lightest $\alpha$-decaying
neutron-deficient polonium isotopes ($A < 196$)  are very different from those of the daughter lead isotopes \cite{And99,Kar06}.
It should be mentioned that our generic form on the evolution of $F(R)$ presented here and in Ref. \cite{And12} is mainly guided by available experimental data. A systematic microscopic calculation on $F(R)$ is desired to confirm this conjecture.

In conclusion, we have studied the origin and physical meaning
of the coefficients $A(Z)$ and $B(Z)$ in the GN law. 
These coefficients are determined from experimental data and show a linear dependence upon $Z$. However, the $Z$-dependence is different in  different regions of the nuclear chart.
Starting
from the microscopic Thomas expression for the decay half life we show that 
$A(Z)$ models the tunneling process as well as the relatively small variations in the 
structure of the neighboring nuclei. The
parameter $B(Z)$ takes into account the clusterization of the $\alpha$-particle
in the mother nucleus. We show why the coefficient $B(Z)$ is negative
and that both $A(Z)$ and $B(Z)$ have to be practically linearly dependent
upon $Z$. 
We also demonstrated here for the first time that, when the dependence of $log_{10}|RF(R)|^2$ on the neutron 
number is not linear or constant, the GN law is broken.  This 
also explains why the GN law works so well in all $\alpha$ emitters known 
today except for the polonium isotopes, as the data within each isotopic chain are so far limited to a region 
where $log_{10}|RF(R)|^2$ behaves linearly with $N$ or is constant. It is only for the polonium isotopic chain that experimental data have been obtained 
over a wide enough range to observe significant deviations from the GN law.
Within the generic description \cite{And12}, the different values of the alpha formation probability for regions I and III and the exponential decrease as a function of neutron number when approaching $N=126$ for region II, can be understood as due to the available $j$ orbitals and a difference in the clustering properties of the nucleons in the $\alpha$ particle. Clustering of the two protons and two neutrons leading to the $\alpha$-particle formation proceeds through high-lying empty single particle configurations. 
 It 
would therefore be very interesting to extend the experimental knowledge towards more 
neutron deficient radon, radium and thorium isotopes in region II and to more neutron-rich lead and mercury isotopes in region I \& IV. This will allow us to validate the generic description, identify the saturation levels of the $\alpha$ formation probability and to investigate the influence of protons and neutrons filling the same single particle orbitals (between 82 and 126). Consequently, compared to the use of the so far generally accepted GN description, more reliable predictions of the $\alpha$ decay half lives will be achieved in unknown nuclei  and in low $\alpha$-decay branching ratios close to stability.

\section*{Acknowledgments}
This work has been supported by the Swedish Research Council (VR) under grant Nos. 621-2012-3805 and
621-2013-4323, FWO-Vlaanderen (Belgium), GOA/2004/03 (BOF-K.U.Leuven), the IUAP-Belgian State-Belgian Science Policy-(BriX network P7/12), and by the European Commission within the Sixth Framework Programme
through I3-ENSAR (contract no. RII3-CT-2010-262010) and by the STFC of UK.

\end{document}